\documentclass[10pt,reqno]{article}
\usepackage{fullpage} 
\usepackage{amssymb} 
\usepackage{graphicx} 
\usepackage{amsmath} 
\usepackage{amsthm,amssymb,latexsym} 
\usepackage{amstext, amsfonts,amsbsy} 
\usepackage{enumerate} 
\usepackage{upgreek} 
\usepackage{float} 
\usepackage{color} 
\usepackage{accents} 
\usepackage{mathtools} 
\usepackage{tikz}
\usetikzlibrary{matrix,graphs,arrows,positioning,calc,decorations.markings,shapes.symbols,shapes.geometric,shapes.misc}
\usepackage{calc, ifthen, ae, xstring, etoolbox, xkeyval,pgfplots}
\usepackage[pdftex,bookmarks,colorlinks,breaklinks]{hyperref}  
\definecolor{dullmagenta}{rgb}{0.4,0,0.4}   
\definecolor{darkblue}{rgb}{0,0,0.4}
\hypersetup{linkcolor=red,citecolor=blue,filecolor=dullmagenta,urlcolor=darkblue} 
\usepackage{eucal}

\newtheorem{theorem}{Theorem}

\newtheorem{lemma}[theorem]{Lemma}

\newcommand{\Pain}[1]{\text{P}_{\mathrm{#1}}}

\newcommand{\Ham}[2]{\mathrm{H}^{\mathrm{#1}}_{\mathrm{#2}}}
\newcommand{\Symp}[1]{\Omega^{\mathrm{#1}}}
\newcommand{\symp}[1]{\omega^{\mathrm{#1}}}

\theoremstyle{definition}

\theoremstyle{remark}

\newtheorem{notation}{Notation}
\newtheorem*{notation*}{Notation} 

\numberwithin{equation}{section}

\begin{document}

{\noindent\Large\bf On some examples of identifying Painlev\'e equations using the geometry of the Okamoto space of initial conditions 
}
\medskip

\begin{flushleft}

\textbf{Anton Dzhamay}\\
Beijing Institute of Mathematical Sciences and Applications (BIMSA)\\ 
No.~544, Hefangkou Village Huaibei Town, Huairou District, Beijing 101408 \\
E-mail: \href{mailto:adzham@bimsa.cn}{\texttt{adzham@bimsa.cn}}\qquad 
ORCID ID: \href{https://orcid.org/0000-0001-8400-2406}{\texttt{0000-0001-8400-2406}}\\[5pt]



\end{flushleft}

%
%
\date{}

\begin{abstract}
	In this short note we give two examples of using the algebro-geometric theory of Painlev\'e equations to solve the Painlev\'e identification problem. The equations 
	that we consider were recently obtained by M.~van der Put and J.~Top in their study of a certain ansatz of isomonodromic deformations of linear ODEs. We provide
	explicit coordinate transformations identifying these examples with standard form of some Painlev\'e equations and also explicitly identify their Hamiltonians.
\end{abstract}


\section{Introduction} 
Differential Painlev\'e equations is an important class of nonlinear second-order differential equations that satisfy the Painlev\'e property that 
a general solution of the equation has no movable, i.e., dependent on the initial conditions, singularities where the solution loses the 
single-valuedness property. These equations were obtained over a hundred years ago in the work of P.~Painlev\'e and his students, especially B.~Gambier
\cite{Pai:1902:EDSODSDLGU}, and almost at the same time in the work of R.~Fuchs on \cite{Fuc:1905:QEDLSO} on isomonodromic deformations. Since then
many important developments in the theory of Painlev\'e equations took place and we recommend the books 
\cite{Con:1999:PP, ConMus:2020:PH, IwaKimShiYos:1991:FGP, FokItsKapNov:2006:PT, GroLaiShi:2002:PDECP}
and references therein for more information.

It is now clear that Painlev\'e equations, and their solution, the so-called \emph{Painlev\'e transcendents}, play an important role in a wide range of important problems in 
Mathematics and Mathematical Physics. However, Painlev\'e equations are not very easy to recognize and to identify -- when expressed in the coordinates
of the applied problem, the equation may look very different from its ``standard'' form. Following \cite{Cla:2019:OPPE}, making such an identification 
is known as the \emph{Painlev\'e equivalence problem}. Matching a Painlev\'e-type equation to its standard form, as well as matching parameters of the problem 
with the Painlev\'e parameters,  immediately allows one to use many known results, such as B\"acklund transformations, existence of special solutions for 
certain parameter values, and so on. Thus, having an effective way to solve the Painlev\'e equivalence problem is very important not just from the theoretical, 
but also from the practical point of view.  

In a recent paper \cite{DzhFilLigSto:2024:DHDPETIUGA} we proposed an essentially algorithmic step-by-step procedure for solving the Painlev\'e equivalence
problem using the algebro-geometric theory of Painlev\'e equations. This geometric approach originated in the work of K.~Okamoto, who introduced the key 
notion of the \emph{space of initial condition}
\cite{Oka:1979:FAESOPCFPP}. This approach was further extended to discrete Painlev\'e equations by H.~Sakai in the seminal paper 
\cite{Sak:2001:RSAWARSGPE}.

The goal of this short paper is to further illustrate the effectiveness of the approach in \cite{DzhFilLigSto:2024:DHDPETIUGA} by applying it to two  examples 
recently obtained by M.~van der Put and J.~Top in their study of a certain ansatz of isomonodromic deformations of linear ODEs \cite{PutTop:2024:IPTECS}. 
The difficulty of the {Painlev\'e equivalence problem}, as well as the effectiveness of the algebro-geometric approach, can be seen from the fact
that in both cases the final identification turned out to be quite different from what the authors, who are well-known experts in this field, initially expected. Our approach
also provides an explicit change of variables transforming the given system into the standard form, as well as constructing the Hamiltonian for the system.

In the nutshell, the algebro-geometric approach for solving the Painlev\'e identification problem for a rational second order ODE or, equivalently, for 
a rational first-order system of two equations, consists of the following steps:
\begin{itemize}
	\item if necessary, rewrite the equation as a first order system and then extend it to $\mathbb{P}^{1} \times \mathbb{P}^{1}$;
	\item identify the \emph{base points} where the numerator and the denominator of rational functions 
	defining the evolution simultaneously vanish and then resolve these points using the \emph{blowup procedure};
	\item each blowup procedure introduces two new charts, extend the dynamics to those charts (effectively, to the exceptional divisor of the blowup) and check for new base points;
	\item once the system is completely regularized, we essentially obtain the \emph{Okamoto space of initial conditions} of the system, we just need to exclude 
	the configuration of \emph{vertical leaves} where the time component of the flow vanishes;
	\item the configuration of {vertical leaves} or, equivalently, the configuration of the \emph{irreducible components of the pole divisor of the symplectic form}, determines
	the \emph{surface type} of the space of initial conditions, which then uniquely identifies the corresponding differential Painlev\'e equation;
	\item comparing the classes of the vertical leaves in the Picard lattice of the resulting surface with those for the \emph{standard geometric realization} allows us
	to construct the matching of two geometries, first on the level a linear change of basis of the Picard lattice, and then by pushing it down to the level of explicit
	birational change of coordinates;
	\item an important role in this process is played by certain canonical parameters, known as the \emph{root variables}, that allow us to match the parameters of the problem
	with the Painlev\'e parameters;
	\item finally, we can use this information to obtain both the symplectic form and the Hamiltonian description of the system and match them to the standard case.
\end{itemize}
This scheme is explained in detail, and illustrated by many examples, in \cite{DzhFilLigSto:2024:DHDPETIUGA}. Thus, here we only provide the outline of the above steps and list the
results of main computations. The main results of the paper are Theorems \ref{thm:coords-vdPT-KNY-6} and  \ref{thm:coords-IP-Ok-4}
that give explicit birational changes of variables making the identifications. 
  
\label{sec:introduction}


\section{Example 1: Painlev\'e $\Pain{VI}$ equation} 
\label{sec:P6}
We first consider the system appearing in \cite[Section 5, pg.~20]{PutTop:2024:IPTECS}. To avoid the clash of variables with the standard notation of the geometric Painlev\'e theory, we rewrite the equations using the variables $x$ for $a_{4}$, $y$ for $a_{1}$, and $s$ in place of $t$. We also use 
$\zeta_{i}$ in place of $P_{i}$ for parameters. Then the system that we want to study becomes
\begin{equation}\label{eq:ex5}
	\left\{
	\begin{aligned}
		\frac{dx}{ds} &= \frac{(-2 x^2 + 2) y s^{2} + 2 (x^{2} + 1)(\zeta_{0} - 2y)s + 2 (x-1)(x+1)(-y + \zeta_{0})}{s^{3} - s},\\
		\frac{dy}{ds} &= \frac{2 \zeta_{1} z^{2} + (- 4 \zeta_{0}^{2} + (8 y + 2)\zeta_{0} - 8 y^2 - 4 \zeta_{2})x + 2 \zeta_{1}}{(x^2 - 1)(s^{2} - 1)}.
	\end{aligned}
	\right.
\end{equation}
It turns out that it is convenient to introduce parameters $\lambda_{1}$ and $\lambda_{2}$ via
\begin{equation}\label{eq:ex5-new-pars}
	\begin{aligned}
		\zeta_{1} &= - \frac{\lambda_{1}^{2} - \lambda_{2}^{2}}{4},\\
		\zeta_{2} &= \frac{\zeta_{0} - \zeta_{0}^{2}}{2} - \frac{\lambda_{1}^{2} + \lambda_{2}^{2}}{4},
	\end{aligned}\qquad\text{or}\qquad
	\begin{aligned}
	\lambda_{1}^{2} &= \zeta_{0} - \zeta_{0}^2 - 2 \zeta_{1} - 2 \zeta_{2},	\\
	\lambda_{2}^{2} &= \zeta_{0} - \zeta_{0}^2 + 2 \zeta_{1} - 2 \zeta_{2}.
	\end{aligned}
\end{equation}
Then \eqref{eq:ex5} becomes
\begin{equation}\label{eq:ex5-mod}
	\left\{
	\begin{aligned}
		\frac{dx}{ds} &= \frac{2s(x^2 + 1)(\zeta_{0} - 2y) - 2 (x^{2}-1)((s^{2}+1)y - \zeta_{0})}{s(s^{2} - 1)},\\
		\frac{dy}{ds} &= \frac{\lambda_{2}^2 (x+1)^2  - \lambda_{1}^{2}(x-1)^{2}-4x (2y - \zeta_{0})^{2}}{(x^2 - 1)(s^{2} - 1)}.
	\end{aligned}
	\right.
\end{equation}

We begin by constructing the Okamoto space of initial conditions for this system by first extending it to $\mathbb{P}^{1} \times \mathbb{P}^{1}$ by introducing the charts $(X,y)$, $(x,Y)$, and $(X,Y)$ via $X = 1/x$, $Y = 1/y$, and then resolving the indeterminacies using the blowup procedure. 
Note that a blowup at a point $q_{i}(x_{i}, y_{i})$ introduces two new coordinate charts $(u_{i},v_{i})$ and $(U_{i},V_{i})$ via
\begin{equation*}
	x = x_{i} + u_{i} = x_{i} + U_{i} V_{i},\quad y = y_{i} + u_{i} v_{i} = y_{i} + V_{i},\qquad u_{i} = x- x_{i}, \quad 
	v_{i} = \frac{y - y_{i}}{x - x_{i}} = \frac{1}{U_{i}},\quad V_{i} = y - y_{i}.
\end{equation*}

The configuration of the indeterminate, or base, points is 
\begin{equation}\label{eq:PT6-pts}
	\begin{aligned}
	&q_{1}\left(x=1, y=\frac{\zeta_{0} - \lambda_{2}}{2}\right),\quad q_{2}\left(x=1, y=\frac{\zeta_{0} + \lambda_{2}}{2}\right),\\
	&q_{3}\left(\frac{1-s}{1+s},\infty\right)\leftarrow 
	q_{4}\left(U_{3} = \left(x-\frac{1-s}{1+s}\right)y = -\frac{s(1 + 2 \zeta_{0})}{(s+1)^{2}}, V_{3} = \frac{1}{y} = 0\right),\\
	&q_{5}\left(x=-1, y=\frac{\zeta_{0} - \lambda_{1}}{2}\right),\quad q_{6}\left(x=-1, y=\frac{\zeta_{0} + \lambda_{1}}{2}\right),\\
	&q_{7}\left(-\frac{1-s}{1+s},\infty\right)\leftarrow 
	q_{8}\left(U_{3} = \left(x+\frac{1-s}{1+s}\right)y = -\frac{s(1 - 2 \zeta_{0})}{(s+1)^{2}}, V_{3} = \frac{1}{y} = 0\right).
	\end{aligned}
\end{equation}

\begin{notation}
In what follows, when points appear in the degeneration cascades, 
\begin{equation*}
	q_{i}\leftarrow q_{j}(u_{i}=0, v_{i} = b_{j} = c_{j}^{-1}) = q_{j}(U_{i} = b_{j}^{-1} = c_{j}, V_{i}=0),	
\end{equation*}
we omit the coordinate system and write $q_{j}(0,b_{j})$ or $q_{j}(c_{j},0)$ unless either $b_{j}$ or $c_{j}$ vanish, then we write everything explicitly.	
\end{notation}

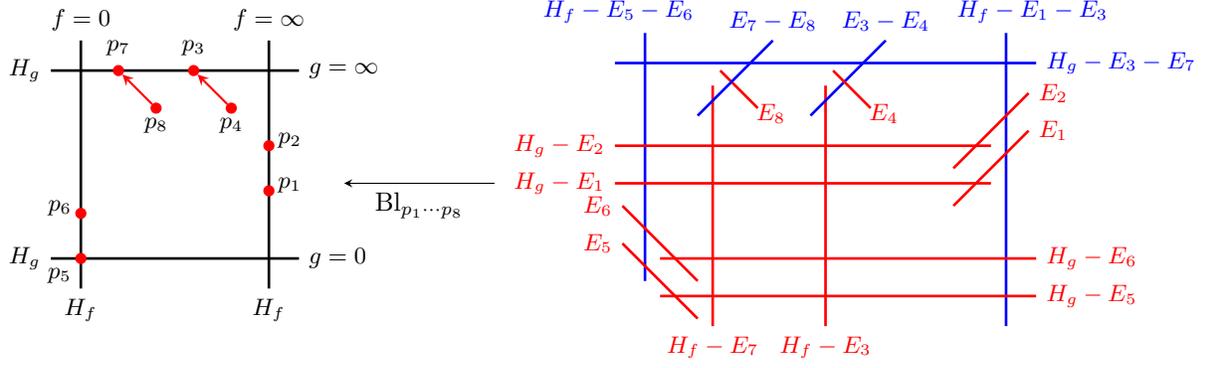
\begin{figure}[ht]
	\begin{tikzpicture}[>=stealth,basept/.style={circle, draw=red!100, fill=red!100, thick, inner sep=0pt,minimum size=1.2mm}]
	\begin{scope}[xshift=0cm,yshift=0cm]
	\draw [black, line width = 1pt] (-0.4,0) -- (2.9,0)	node [pos=0,left] {\small $H_{g}$} node [pos=1,right] {\small $g=0$};
	\draw [black, line width = 1pt] (-0.4,2.5) -- (2.9,2.5) node [pos=0,left] {\small $H_{g}$} node [pos=1,right] {\small $g=\infty$};
	\draw [black, line width = 1pt] (0,-0.4) -- (0,2.9) node [pos=0,below] {\small $H_{f}$} node [pos=1,above] {\small $f=0$};
	\draw [black, line width = 1pt] (2.5,-0.4) -- (2.5,2.9) node [pos=0,below] {\small $H_{f}$} node [pos=1,above] {\small $f=\infty$};
	\node (p1) at (2.5,0.9) [basept,label={[right] \small $p_{1}$}] {};
	\node (p2) at (2.5,1.5) [basept,label={[right] \small $p_{2}$}] {};
	\node (p3) at (1.5,2.5) [basept,label={[above] \small $p_{3}$}] {};
	\node (p4) at (2,2) [basept,label={[xshift = 0pt, yshift=-15pt] \small $p_{4}$}] {};
	\node (p5) at (0,0) [basept,label={[xshift = -8pt, yshift=-15pt] \small $p_{5}$}] {};
	\node (p6) at (0,0.6) [basept,label={[left] \small $p_{6}$}] {};
	\node (p7) at (0.5,2.5) [basept,label={[above] \small $p_{7}$}] {};
	\node (p8) at (1,2) [basept,label={[xshift = 0pt, yshift=-15pt] \small $p_{8}$}] {};
	\draw [red, line width = 0.8pt, ->] (p4) -- (p3);
	\draw [red, line width = 0.8pt, ->] (p8) -- (p7);
	\end{scope}
	\draw [->] (5.5,1)--(3.5,1) node[pos=0.5, below] {$\operatorname{Bl}_{p_{1}\cdots p_{8}}$};
	\begin{scope}[xshift=7.5cm,yshift=0cm]
	\draw [blue, line width = 1pt] (0,-0.3) -- (0,3) node [pos=1, above, xshift=-10pt] {\small $H_{f}-E_{5}-E_{6}$};
	\draw [blue, line width = 1pt] (-0.4,2.6) -- (5.2,2.6) node [pos=1,right] {\small $H_{g} - E_{3} - E_{7}$};
	\draw [blue, line width = 1pt] (4.8,-0.9) -- (4.8,3) node [pos=1, above, xshift=10pt] {\small $H_{f}-E_{1} - E_{3}$};	
	\draw [red, line width = 1pt] (0.2,-0.5) -- (5.2,-0.5)	node [pos=1, right] {\small $H_{g}-E_{5}$};
	\draw [red, line width = 1pt] (0.7,-0.8) -- (-0.3,0.2) node [pos=1,left] {\small $E_{5}$};
	\draw [red, line width = 1pt] (0.2,0) -- (5.2,0)	node [pos=1, right] {\small $H_{g}-E_{6}$};
	\draw [red, line width = 1pt] (0.7,-0.3) -- (-0.3,0.7) node [pos=1,left] {\small $E_{6}$};
	\draw [red, line width = 1pt] (-0.4,1) -- (4.6,1)	node [pos=0, left] {\small $H_{g}-E_{1}$};
	\draw [red, line width = 1pt] (4.1,0.7) -- (5.1,1.7) node [pos=1,right] {\small $E_{1}$};
	\draw [red, line width = 1pt] (-0.4,1.5) -- (4.6,1.5)	node [pos=0, left] {\small $H_{g}-E_{2}$};
	\draw [red, line width = 1pt] (4.1,1.2) -- (5.1,2.2) node [pos=1,right] {\small $E_{2}$};
	\draw [red, line width = 1pt] (0.9,-0.9) -- (0.9,2.3) node [pos=0,below] {\small $H_{f} - E_{7}$};
	\draw [red, line width = 1pt] (1,2.5) -- (1.5,2) node [pos=1,xshift=5pt,yshift=-2pt] {\small $E_{8}$};
	\draw [blue, line width = 1pt] (0.7,1.9) -- (1.7,2.9) node [pos=1,above] {\small $E_{7}-E_{8}$};
	\draw [red, line width = 1pt] (2.4,-0.9) -- (2.4,2.3) node [pos=0,below] {\small $H_{f} - E_{3}$};
	\draw [blue, line width = 1pt] (2.2,1.9) -- (3.2,2.9) node [pos=1,above] {\small $E_{3}-E_{4}$};
	\draw [red, line width = 1pt] (2.5,2.5) -- (3,2) node [pos=1,xshift=5pt,yshift=-2pt] {\small $E_{4}$};
	\end{scope}
	\end{tikzpicture}
	\caption{The standard realization of the $D_{4}^{(1)}$ Sakai surface}
	\label{fig:soic-KNY-P6}
\end{figure}

This point configuration is, up to some M\"obius transformations of the coordinates, the same as the standard geometric realization 
of the $D_{4}^{(1)}$ surface family in \cite[Section 8.2.17]{KajNouYam:2017:GAPE} shown on Figure~\ref{fig:soic-KNY-P6}, see also 
\cite[Section 6]{DzhFilLigSto:2024:DHDPETIUGA}\footnote{Note that the roles of the coordinates $(q,p)$ and $(f,g)$ are switched between these two references. Here we follow \cite{KajNouYam:2017:GAPE}.}. There the configuration of the base points is given in coordinates $(f,g)$
in terms of the \emph{root variables} $a_{i}$,
\begin{equation*}
	p_{1}(\infty,-a_{2}),\quad p_{2}(\infty,-a_{1}-a_{2}),\quad p_{3}(t,\infty)\leftarrow p_{4}(ta_{0},0),\quad 
	p_{5}(0,0),\quad p_{6}(0,a_{4}),\quad p_{7}(1,\infty)\leftarrow p_{8}(a_{3},0).
\end{equation*}

%

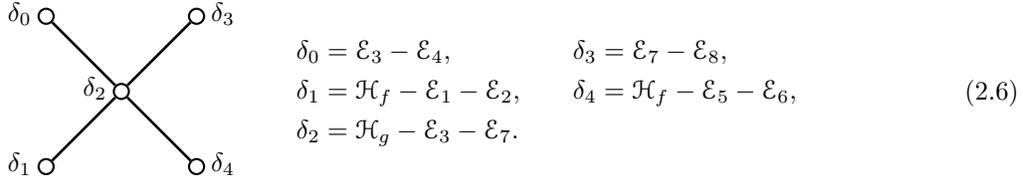
\begin{figure}[ht]
\begin{equation}\label{eq:d-roots-d41}			
	\raisebox{-32.1pt}{\begin{tikzpicture}[
			elt/.style={circle,draw=black!100,thick, inner sep=0pt,minimum size=2mm}]
		\path 	(-1,1) 	node 	(d0) [elt, label={[xshift=-10pt, yshift = -10 pt] $\delta_{0}$} ] {}
		        (-1,-1) node 	(d1) [elt, label={[xshift=-10pt, yshift = -10 pt] $\delta_{1}$} ] {}
		        ( 0,0) 	node  	(d2) [elt, label={[xshift=-10pt, yshift = -10 pt] $\delta_{2}$} ] {}
		        ( 1,1) 	node  	(d3) [elt, label={[xshift=10pt, yshift = -10 pt] $\delta_{3}$} ] {}
		        ( 1,-1) node 	(d4) [elt, label={[xshift=10pt, yshift = -10 pt] $\delta_{4}$} ] {};
		\draw [black,line width=1pt ] (d0) -- (d2) -- (d1)  (d3) -- (d2) -- (d4);
	\end{tikzpicture}} \qquad
			\begin{alignedat}{2}
			\delta_{0} &= \mathcal{E}_{3} - \mathcal{E}_{4}, &\qquad  \delta_{3} &= \mathcal{E}_{7} - \mathcal{E}_{8},\\
			\delta_{1} &= \mathcal{H}_{f} - \mathcal{E}_{1} - \mathcal{E}_{2}, &\qquad  \delta_{4} &= \mathcal{H}_{f} - \mathcal{E}_{5} - \mathcal{E}_{6},\\
			\delta_{2} &= \mathcal{H}_{g} - \mathcal{E}_{3} - \mathcal{E}_{7}.
			\end{alignedat}
\end{equation}
	\caption{The Surface Root Basis for the standard $D_{4}^{(1)}$ Sakai surface point configuration}
	\label{fig:d-roots-d41-KNY}	
\end{figure}

The key algebraic object associated to this surface family is its Picard lattice 
\begin{equation}
	\operatorname{Pic}(\mathcal{X}) = \operatorname{Span}\{\mathcal{H}_{1}, \mathcal{H}_{2}, \mathcal{E}_{1},\ldots \mathcal{E}_{8}\}
\end{equation}
and its surface and symmetry sub-lattices, see \cite{DzhFilLigSto:2024:DHDPETIUGA, KajNouYam:2017:GAPE}.
The surface and symmetry root bases for this standard realization of the $D_{4}^{(1)}$ surface are given on Fig.~\ref{fig:d-roots-d41-KNY}	
and Fig.~\ref{fig:a-roots-d41-KNY} respectively. 
\begin{figure}[H]
\begin{equation}\label{eq:a-roots-d41}			
	\raisebox{-32.1pt}{\begin{tikzpicture}[
			elt/.style={circle,draw=black!100,thick, inner sep=0pt,minimum size=2mm}]
		\path 	(-1,1) 	node 	(d0) [elt, label={[xshift=-10pt, yshift = -10 pt] $\alpha_{0}$} ] {}
		        (-1,-1) node 	(d1) [elt, label={[xshift=-10pt, yshift = -10 pt] $\alpha_{1}$} ] {}
		        ( 0,0) 	node  	(d2) [elt, label={[xshift=-10pt, yshift = -10 pt] $\alpha_{2}$} ] {}
		        ( 1,1) 	node  	(d3) [elt, label={[xshift=10pt, yshift = -10 pt] $\alpha_{3}$} ] {}
		        ( 1,-1) node 	(d4) [elt, label={[xshift=10pt, yshift = -10 pt] $\alpha_{4}$} ] {};
		\draw [black,line width=1pt ] (d0) -- (d2) -- (d1)  (d3) -- (d2) -- (d4);
	\end{tikzpicture}} \qquad
			\begin{alignedat}{2}
			\alpha_{0} &=  \mathcal{H}_{f} -\mathcal{E}_{3} - \mathcal{E}_{4}, &\qquad  \alpha_{3} &= \mathcal{H}_{f} - \mathcal{E}_{7} - \mathcal{E}_{8},\\
			\alpha_{1} &= \mathcal{E}_{1} - \mathcal{E}_{2}, &\qquad  \alpha_{4} &= \mathcal{E}_{5} - \mathcal{E}_{6}.\\
			\alpha_{2} &= \mathcal{H}_{g} - \mathcal{E}_{1} - \mathcal{E}_{5},
			\end{alignedat}
\end{equation}
	\caption{The Symmetry Root Basis for the standard $D_{4}^{(1)}$ symmetry sub-lattice}
	\label{fig:a-roots-d41-KNY}	
\end{figure}

From \eqref{eq:PT6-pts} it is clear that points $q_{i}$ lie on the polar divisor of the symplectic form $\omega^{vdPT} = k \frac{dx \wedge dy}{x^{2} - 1}$.
Computing the root variables for this point configuration (see \cite{DzhFilLigSto:2024:DHDPETIUGA} or \cite{DzhTak:2018:SASGTDPE} for details on how to perform such computations), and imposing the normalization condition $a_{0} + a_{2} + 2a_{2} + a_{3} + a_{4} = 1$ we see that $k = -2$ and the root
variables are
\begin{equation}\label{eq:vdPT-root-vars}
	a_{0} = \frac{1}{2} + \zeta_{0} ,\quad a_{1} = \lambda_{2},\quad a_{2} = \frac{\lambda_{1} - \lambda_{2}}{2},\quad a_{3} = \frac{1}{2} - \zeta_{0},\quad 
	a_{4} = - \lambda_{1}.
\end{equation}
Moreover, knowing the symplectic structure suggests how to look for the Hamiltonian for system \eqref{eq:ex5-mod} --- we want the system to have the form
\begin{equation}
	\left\{\begin{aligned}
		\frac{dx}{ds} &= \frac{x^{2}-1}{2}  \frac{\partial \Ham{vdPT}{VI}}{\partial y} 
			= \frac{2s(x^2 + 1)(\zeta_{0} - 2y) - 2 (x^{2}-1)((s^{2}+1)y - \zeta_{0})}{s(s^{2} - 1)},\\
		\frac{dy}{ds} &= -\frac{x^{2}-1}{2}  \frac{\partial \Ham{vdPT}{VI}}{\partial x} = 
				\frac{\lambda_{2}^2 (x+1)^2  - \lambda_{1}^{2}(x-1)^{2}-4x (2y - \zeta_{0})^{2}}{(x^2 - 1)(s^{2} - 1)}.
		\end{aligned}\right.
\end{equation} 
A straightforward direct computation then gives
\begin{equation}\label{eq:vdPT-Ham6}
\Ham{vdPT}{VI}(x,y,s) = \frac{8y(\zeta_{0} - y)}{(s^{2} -1)(x^{2} -1)} + 
	\frac{\lambda_{2}^{2}(x+1) - \lambda_{1}^{2}(x-1)- 2 \zeta_{0}^{2}}{(s^{2}-1)(x^{2}-1)} + 
	\frac{2y(2 \zeta_{0} - (s+1)y)}{s(s-1)}.
\end{equation} 

It is well-known that $D_{4}^{(1)}$ surface family is the space of initial conditions for $\Pain{VI}$ equation, \cite{Oka:1987:SPEISPEP,Sak:2001:RSAWARSGPE},
\begin{equation}\label{eq:P6-std}
	\begin{aligned}
	\frac{d^{2} w}{dt^2} &= \frac{1}{2}\left(\frac{1}{w} + \frac{1}{w-1} + \frac{1}{w-t}\right)\left(\frac{dw}{dt}\right)^{2}
	-\left(\frac{1}{t} + \frac{1}{t-1} + \frac{1}{w-t}\right)\frac{dw}{dt} + \\
	&\qquad \frac{w(w-1)(w-t)}{t^{2}(t-1)^{2}}\left(\alpha + \beta \frac{t}{w^{2}} + \gamma \frac{t-1}{(w-1)^{2}} + 
	\delta\frac{t(t-1)}{(w-t)^{2}}\right).
	\end{aligned}
\end{equation}
Different geometric realizations of this family lead to different Hamiltonian forms of this equation, as explained in detail in 
\cite[Section 6]{DzhFilLigSto:2024:DHDPETIUGA}. The one corresponding to our geometric model is the Kajiwara-Noumi-Yamada Hamiltonian
\begin{equation}\label{eq:KNY-Ham6}
	\Ham{KNY}{VI}(f,g;t) = \frac{(f-1)(f-t)g}{t(t-1)}
	\left(\frac{g}{f}-\left(\frac{a_{0}-1}{f-t} + \frac{a_{3}}{f-1} + \frac{a_{4}}{f}\right)\right) + \frac{a_{2}(a_{1}+a_{2})(f-t)}{t(t-1)}.
\end{equation}
In these coordinates the symplectic form becomes logarithmic, $\symp{KNY}=(1/f)dg\wedge df$ and the resulting Hamiltonian system is 
\begin{equation}\label{eq:KNY-Ham6-sys}
\left\{
\begin{aligned}
	\frac{df}{dt} &= f \frac{\partial \Ham{KNY}{VI}}{\partial g} 
		=\frac{1}{t(t-1)}\Big( (f-1)(f-t)(2g-a_{4}) - (a_{0}-1)f(f-1) -a_{3} f(f-t) \Big),\\
	\frac{dg}{dt} &= -f\frac{\partial \Ham{KNY}{VI}}{\partial f} 
		=-\frac{1}{t(t-1)}\Big( f(g+a_{2})(g+a_{1}+a_{2}) - \frac{t g}{f}(g-a_{4})\Big),
\end{aligned}
\right.
\end{equation}
which reduces to the standard Painlev\'e $\Pain{VI}$ equation \eqref{eq:P6-std} for $f(t)$ with parameters
\begin{equation}\label{eq:pars-root-P6}
	\alpha = \frac{a_{1}^{2}}{2},\quad \beta=-\frac{a_{4}^{2}}{2},\quad \gamma=\frac{a_{3}^{2}}{2},\quad \delta=\frac{1-a_{0}^{2}}{2}.
\end{equation}

To reduce the point configuration $\{q_{i}\}$ in coordinates $(x,y)$ to the standard point configuration $\{p_{i}\}$ in coordinates $(f,g)$ we need to map 
$x=1$ to $f=\infty$, $x=-1$ to $f=0$, $y=\frac{\zeta_{0} - \lambda_{1}}{2}$ to $g=0$, and adjust the scaling on the axes to match \eqref{eq:vdPT-root-vars}. This can be done by M\"obius transformation of the coordinates described in the next Theorem. As a consequence we also obtain the relationship between 
the time variables $s$ and $t$ by matching the coordinates of points $q_{3}$ and $p_{3}$.

\begin{theorem}\label{thm:coords-vdPT-KNY-6} The change of coordinates and parameter matching between system \eqref{eq:ex5-mod} and the Kajiwara-Noumi-Yamada 
	Hamiltonian system \eqref{eq:KNY-Ham6-sys} is given by
    \begin{equation}\label{eq:vdPT-Ok-6}
   	 \left\{\begin{aligned}
   	 	f(x,y,t)&=\frac{1+x}{s(1-x)},\\
   		g(x,y,t)&= -y + \frac{\zeta_{0} - \lambda_{1}}{2},\\
		t(s) &= \frac{1}{s^{2}},
   	 \end{aligned}\right.
    \qquad\text{and} \quad 
    	\left\{\begin{aligned}
   	 	x(f,g,t)&=\frac{fs -1}{fs +1},\\
   		y(f,g,t)&= -g + \frac{\zeta_{0} - \lambda_{1}}{2}, &
		s(t) &= \frac{1}{\sqrt{t}},
    	\end{aligned}\right.
    \end{equation}
	with the parameter correspondence given by \eqref{eq:vdPT-root-vars}.
\end{theorem}

Direct computation then shows that a solution of \eqref{eq:ex5} gives a solution $w(t) = \frac{1 + x(1/\sqrt{t})}{\sqrt{t}(1-x(1/\sqrt{t}))}$ of the 
standard $\Pain{VI}$ equation \eqref{eq:P6-std} with parameters
\begin{equation}
	\alpha = \frac{\lambda_{2}^{2}}{2} = \frac{\zeta_{0}(1-\zeta_{0})}{2} + \zeta_{1} - \zeta_{2},\quad 
	\beta = - \frac{\lambda_{1}^{2}}{2} = \frac{\zeta_{0}(\zeta_{0}-1)}{2} + \zeta_{1} + \zeta_{2},\quad
	\gamma = \frac{(1 - 2 \zeta_{0})^{2}}{8},\quad \delta = \frac{3-4\zeta_{0}(1 + \zeta_{0})}{8}.
\end{equation}

Another well-known Hamiltonian form of $\Pain{VI}$ equation is due to Okamoto \cite{Oka:1987:SPEIFPEP} (and Malmquist \cite{Mal:1922:EDSODLGPCF}).
For this example the geometry is slightly more complicated (the space of the initial conditions is not minimal), but the symplectic form 
is the standard one, $\symp{Ok} = dp\wedge dq$. The Hamiltonian is
\begin{equation}\label{eq:Ok-Ham6}
	\Ham{Ok}{VI}(q,p;t) = \frac{1}{t(t-1)}\Big(q(q-1)(q-t)p^{2} - \big(
	\kappa_{0}(q-1)(q-t) + \kappa_{1}q(q-t)+ (\theta-1)q(q-1)\big)p + \kappa (q-t)\Big)
\end{equation}
and the resulting system is 
  \begin{equation}\label{eq:Ok-Ham6-sys}
  	\left\{
 	\begin{aligned}
 		\frac{dq}{dt} &= \frac{\partial \Ham{Ok}{VI}}{\partial p} 
			=\frac{1}{t(t-1)}\Big( (q-1)(q-t)(2 q p - \kappa_{0}) -\kappa_{1} q(q-t) - (\theta-1) q (q-1)\Big),\\
		\frac{dp}{dt} &= -\frac{\partial \Ham{Ok}{VI}}{\partial q} 
			=-\frac{1}{t(t-1)}\Big( \big( (3q^2 -2(t+1)q +t)p - (\kappa_{0} + \kappa_{1}) (2q-t) - (\theta-1)(2q-1) + \kappa_{0}\big)p + \kappa\Big).
 	\end{aligned}
	\right.
  \end{equation}
Here the \emph{Okamoto parameters}
$\kappa_{0}$, $\kappa_{1}$, $\kappa_{\infty}$, and $\theta$ are given by
\begin{equation}\label{eq:Ok-pars-PVI}
	\alpha = \frac{\kappa_{\infty}^{2}}{2},\quad \beta = -\frac{\kappa_{0}^{2}}{2},\quad \gamma = \frac{\kappa_{1}^{2}}{2}, 
		\quad \delta = \frac{1-\theta^{2}}{2},
	\quad\text{as well as } \kappa = \frac{(\kappa_{0}+\kappa_{1}+\theta-1)^{2} - \kappa_{\infty}^{2}}{4},
\end{equation}
or, in terms of the root variables, 
\begin{equation}\label{eq:pars-Ok2root-P6}
	\theta = a_{0}, \quad \kappa_{0} = a_{4}, \quad \kappa_{1} = a_{3}, \quad \kappa_{\infty} = - a_{1}.
\end{equation}
These two models are related by the change of variables $(f,g) = (q,qp)$. 

Given that the change of coordinates is time-dependent, 
to compare the Hamiltonians for different coordinate systems we use the equality 
of pull-backs of the following $2$-form on the extended phase space. We use the change of 
coordinates $\varphi:(x,y,s)\to (q,p,t) = \left(\frac{1+x}{s(1-x)},\frac{s(1-x)(- 2y + \zeta_{0} - \lambda_{1})}{2(1 + x)}, \frac{1}{s^{2}}\right)$ 
to match the Hamiltonian \eqref{eq:vdPT-Ham6} and the Okamoto Hamiltonian \eqref{eq:Ok-Ham6}:
\begin{align*}
	\Symp{vdPT} &= \symp{vdPT} - d\Ham{vdPT}{VI}\wedge ds = \varphi^{*}(\Symp{Ok}) =  \varphi^{*}\left(\symp{Ok} - d\Ham{Ok}{VI}\wedge dt\right) \\
	&= \frac{\partial p}{\partial y}\frac{\partial q}{\partial x} dy\wedge dx + 
	\left(\frac{\partial p}{\partial x}\frac{\partial q}{\partial s} - \frac{\partial p}{\partial s}\frac{\partial q}{\partial x}\right) dx\wedge dx 
	+ \frac{\partial p}{\partial y}\frac{\partial q}{\partial s} dy\wedge ds - d\Ham{Ok}{VI}\wedge \left(\frac{-2}{s^{3}}ds\right)\\
	&= \frac{2}{x^{2} - 1} dy \wedge dx + \frac{1}{s} dy \wedge ds - d\left(\frac{-2}{s^{3}}\Ham{Ok}{VI}(q(x,y,s), p(x,y,s), t(s)) \right) \wedge ds.
\end{align*}
Thus we see that 
\begin{equation*}
	\Ham{vdPT}{VI}(x,y,s) = \frac{-2}{s^{3}}\Ham{Ok}{VI}\left(q(x,y,s),p(x,y,s),t(s)\right) - \frac{y}{s} + h(s),
\end{equation*}
where the last term is purely time-dependent and does not effect the equations of motion. A direct computation shows that 
\begin{equation}
	h(s) =  - \frac{(s-3)\zeta_{0}^{2} - (s-1)\zeta_{0}(2 \lambda_{1} + 1) + \lambda_{1} (s \lambda_{1} + s - 1) + \lambda_{2}^{2}}{2 s (s-1)}.
\end{equation}

\section{Example 2: Painlev\'e $\Pain{IV}$ equation} 
\label{sec:P4}
As a second example we consider the Hamiltonian system appearing in \cite[Section 9, pg.~26]{PutTop:2024:IPTECS}. We again rewrite the equations using the variables $x$ for $a_{4}$, $y$ for $a_{2}$, $s$ in place of $t$, and $\xi_{i}$ for parameters $p_{i}$. The Hamiltonian then takes the form 
\begin{equation}\label{eq:ex9-Ham}
	H(x,y,s; \xi_{i}) = - 3 x^{2} y^{3} - \frac{9 \xi_{1}- 12 \xi_{2}}{2} x y^{2} - \frac{3\xi_{1}^{2} - 9 \xi_{1} \xi_{2} + 6 \xi_{2}^{2}}{2} y
		+ \frac{3}{2} x + \frac{9 x y}{2}s,
\end{equation}
and the equations of motion, with respect to the standard symplectic structure $dy\wedge dx$, are
\begin{equation}\label{eq:ex9}
	\left\{
	\begin{aligned}
		\frac{dx}{ds} &= \frac{\partial H}{\partial y} = \frac{3(-\xi_{1} + 2 \xi_{2})(-\xi_{2} + \xi_{1})}{2} + \frac{9 s}{2} x
			+ (-9\xi_{1} + 12\xi_{2}) x y - 9 x^{2} y^{2},\\
		\frac{dy}{ds} &= - \frac{\partial H}{\partial x} = -\frac{3}{2} + 6 x y^{3} + \frac{9 \xi_{1} - 12 \xi_{2}}{2} y^{2} - \frac{9 y s}{2}.
	\end{aligned}
	\right.
\end{equation}

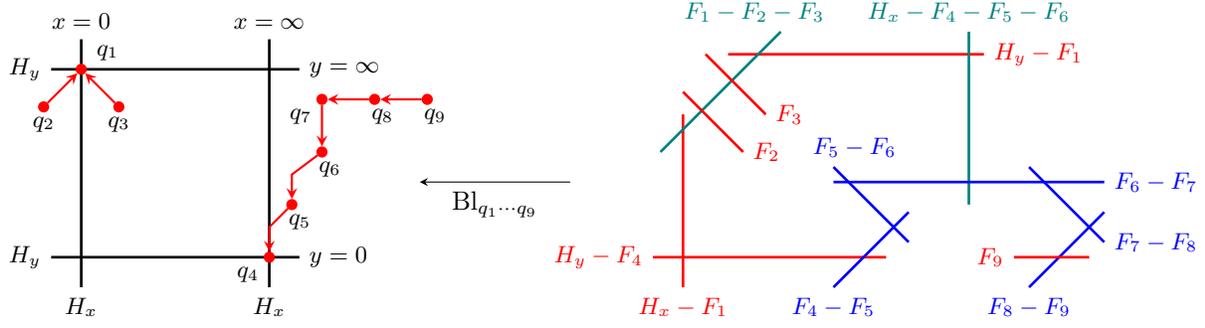
\begin{figure}[ht]
	\begin{tikzpicture}[>=stealth,basept/.style={circle, draw=red!100, fill=red!100, thick, inner sep=0pt,minimum size=1.2mm}]
	\begin{scope}[xshift=0cm,yshift=0cm]
	\draw [black, line width = 1pt] (-0.4,0) -- (2.9,0)	node [pos=0,left] {\small $H_{y}$} node [pos=1,right] {\small $y=0$};
	\draw [black, line width = 1pt] (-0.4,2.5) -- (2.9,2.5) node [pos=0,left] {\small $H_{y}$} node [pos=1,right] {\small $y=\infty$};
	\draw [black, line width = 1pt] (0,-0.4) -- (0,2.9) node [pos=0,below] {\small $H_{x}$} node [pos=1,above] {\small $x=0$};
	\draw [black, line width = 1pt] (2.5,-0.4) -- (2.5,2.9) node [pos=0,below] {\small $H_{x}$} node [pos=1,above] {\small $x=\infty$};
	\node (q1) at (0,2.5) [basept,label={[xshift = 10pt, yshift=-2pt] \small $q_{1}$}] {};
	\node (q2) at (-0.5,2) [basept,label={[xshift = 0pt, yshift=-15pt] \small $q_{2}$}] {};
	\node (q3) at (0.5,2) [basept,label={[xshift = 0pt, yshift=-15pt] \small $q_{3}$}] {};
	\node (q4) at (2.5,0) [basept,label={[xshift = -8pt, yshift=-15pt] \small $q_{4}$}] {};
	\node (q5) at (2.8,0.7) [basept,label={[xshift = 3pt, yshift=-15pt] \small $q_{5}$}] {};
	\node (q6) at (3.2,1.4) [basept,label={[xshift = 3pt,yshift=-15pt] \small $q_{6}$}] {};
	\node (q7) at (3.2,2.1) [basept,label={[xshift = -8pt,yshift=-15pt] \small $q_{7}$}] {};
	\node (q8) at (3.9,2.1) [basept,label={[xshift = 3pt,yshift=-15pt] \small $q_{8}$}] {};
	\node (q9) at (4.6,2.1) [basept,label={[xshift = 3pt,yshift=-15pt] \small $q_{9}$}] {};
	\draw [red, line width = 0.8pt, ->] (q2) -- (q1);
	\draw [red, line width = 0.8pt, ->] (q3) -- (q1);
	\draw [red, line width = 0.8pt, ->] (q5) -- (2.5,0.4) -- (q4);
	\draw [red, line width = 0.8pt, ->] (q6) -- (2.8,1.1) -- (q5);
	\draw [red, line width = 0.8pt, ->] (q7) -- (q6);
	\draw [red, line width = 0.8pt, ->] (q8) -- (q7);	
	\draw [red, line width = 0.8pt, ->] (q9) -- (q8);	
	\end{scope}
	\draw [->] (6.5,1)--(4.5,1) node[pos=0.5, below] {$\operatorname{Bl}_{q_{1}\cdots q_{9}}$};
	\begin{scope}[xshift=8cm,yshift=0cm]
	\draw [red, line width = 1pt] (0,-0.4) -- (0,1.9) node [pos=0, below] {\small $H_{x}-F_{1}$};
	\draw [red, line width = 1pt] (0.6,2.7) -- (4,2.7) node [pos=1,right] {\small $H_{y} - F_{1}$};
	\draw [teal, line width = 1pt] (-0.3,1.4) -- (1.3,3) node [pos=1,pos=1,above, xshift=-10pt] {\small $F_{1} - F_{2} - F_{3}$};
	\draw [red, line width = 1pt] (0,2.2) -- (0.8,1.4) node [pos=1, right] {\small $F_{2}$};
	\draw [red, line width = 1pt] (0.3,2.7) -- (1.1,1.9) node [pos=1, right] {\small $F_{3}$};
	\draw [teal, line width = 1pt] (3.8,0.7) -- (3.8,3) node [pos=1, above] {\small $H_{x}-F_{4}-F_{5}-F_{6}$};
	\draw [red, line width = 1pt] (-0.4,0) -- (2.7,0)	node [pos=0, left] {\small $H_{y}-F_{4}$};
	\draw [blue, line width = 1pt] (2,-0.4) -- (3,0.6) node [pos=0, below] {\small $F_{4}-F_{5}$};
	\draw [blue, line width = 1pt] (3,0.2) -- (2,1.2) node [pos=1, above, xshift=8pt] {\small $F_{5}-F_{6}$};
	\draw [blue, line width = 1pt] (2,1) -- (5.6,1)	node [pos=1, right] {\small $F_{6}-F_{7}$};
	\draw [blue, line width = 1pt] (5.6,0.2) -- (4.6,1.2) node [pos=0, right] {\small $F_{7}-F_{8}$};
	\draw [blue, line width = 1pt] (5.6,0.6) -- (4.6,-0.4) node [pos=1, below] {\small $F_{8}-F_{9}$};
	\draw [red, line width = 1pt] (4.4,0) -- (5.4,0)	node [pos=0, left] {\small $F_{9}$};
	\end{scope}
	\end{tikzpicture}
	\caption{The Space of Initial Conditions for the van der Put-Top  Hamiltonian System}
	\label{fig:vdPT-soic-4}
\end{figure}

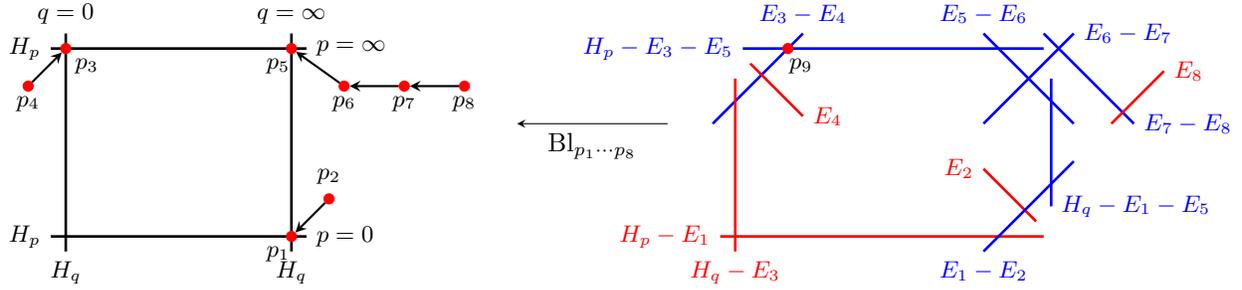
\begin{figure}[ht]
	\begin{tikzpicture}[>=stealth,basept/.style={circle, draw=red!100, fill=red!100, thick, inner sep=0pt,minimum size=1.2mm}]
	\begin{scope}[xshift=0cm,yshift=0cm]
	\draw [black, line width = 1pt] (-0.2,0) -- (3.2,0)	node [pos=0,left] {\small $H_{p}$} node [pos=1,right] {\small $p=0$};
	\draw [black, line width = 1pt] (-0.2,2.5) -- (3.2,2.5) node [pos=0,left] {\small $H_{p}$} node [pos=1,right] {\small $p=\infty$};
	\draw [black, line width = 1pt] (0,-0.2) -- (0,2.7) node [pos=0,below] {\small $H_{q}$} node [pos=1,above] {\small $q=0$};
	\draw [black, line width = 1pt] (3,-0.2) -- (3,2.7) node [pos=0,below] {\small $H_{q}$} node [pos=1,above] {\small $q=\infty$};
	\node (p1) at (3,0) [basept,label={[xshift = -5pt, yshift=-15pt] \small $p_{1}$}] {};
	\node (p2) at (3.5,0.5) [basept,label={[yshift=0pt] \small $p_{2}$}] {};
	\node (p3) at (0,2.5) [basept,label={[xshift = 8pt, yshift=-15pt] \small $p_{3}$}] {};
	\node (p4) at (-0.5,2) [basept,label={[yshift=-15pt] \small $p_{4}$}] {};
	\node (p5) at (3,2.5) [basept,label={[xshift = -5pt, yshift=-15pt] \small $p_{5}$}] {};
	\node (p6) at (3.7,2) [basept,label={[yshift=-15pt] \small $p_{6}$}] {};
	\node (p7) at (4.5,2) [basept,label={[yshift=-15pt] \small $p_{7}$}] {};
	\node (p8) at (5.3,2) [basept,label={[yshift=-15pt] \small $p_{8}$}] {};
	\draw [line width = 0.8pt, ->] (p2) -- (p1);
	\draw [line width = 0.8pt, ->] (p4) -- (p3);	
	\draw [line width = 0.8pt, ->] (p6) -- (p5);
	\draw [line width = 0.8pt, ->] (p7) -- (p6);
	\draw [line width = 0.8pt, ->] (p8) -- (p7);
	\end{scope}
	\draw [->] (8,1.5)--(6,1.5) node[pos=0.5, below] {$\operatorname{Bl}_{p_{1}\cdots p_{8}}$};
	\begin{scope}[xshift=9.5cm,yshift=0cm]
	\draw [blue, line width = 1pt] (-0.5,2.5) -- (3.5,2.5) node [pos=0,left] {\small $H_{p} - E_{3} - E_{5}$};
	\draw [blue, line width = 1pt] (-0.9,1.5) -- (0.3,2.7) node [pos=1,above] {\small $E_{3} - E_{4}$};		
		\node (p9) at (0.1,2.5) [basept,label={[xshift = 5pt, yshift=-15pt] \small $p_{9}$}] {};
	\draw [blue, line width = 1pt] (3.9,1.5) -- (2.7,2.7) node [pos=1,above] {\small $E_{5} - E_{6}$};		
	\draw [red, line width = 1pt] (-0.8,0) -- (3.5,0)	node [pos=0,left] {\small $H_{p}-E_{1}$};
	\draw [red, line width = 1pt] (-0.6,-0.2) -- (-0.6,2.1) node [pos=0,below] {\small $H_{q} - E_{3}$};
	\draw [red, line width = 1pt] (-0.4,2.3) -- (0.3,1.6)	node [pos=1,right] {\small $E_{4}$};
	\draw [red, line width = 1pt] (3.4,0.2) -- (2.7,0.9)	node [pos=1,left] {\small $E_{2}$};
	\draw [blue, line width = 1pt] (2.7,1.5) -- (3.9,2.7)	node [pos=1,right] {\small $E_{6}-E_{7}$};
	\draw [blue, line width = 1pt] (4.7,1.5) -- (3.5,2.7) node [pos=0,right] {\small $E_{7} - E_{8}$};		
	\draw [blue, line width = 1pt] (3.9,1) -- (2.7,-0.2) node [pos=1,below] {\small $E_{1} - E_{2}$};		
	\draw [blue, line width = 1pt] (3.6,0.4) -- (3.6,2.1) node [pos=0,right] {\small $H_{q} - E_{1} - E_{5}$};
	\draw [red, line width = 1pt] (4.4,1.5) -- (5.1,2.2) node [pos=1,right] {\small $E_{8}$};
	\end{scope}
	\end{tikzpicture}
	\caption{The standard $E_{6}^{(1)}$ Sakai surface (with an additional blow-up point)}
	\label{fig:surface-e61}
\end{figure}

This system has \emph{nine} base points that come in two degeneration cascades,
\begin{equation}\label{eq:Ok-pts-5}
	\begin{tikzpicture}[baseline=0pt]
	\node (q1) at (0,0) {$q_{1}(0,\infty)$}; 
	\node (q2) at (2.5,0.7) {$q_{2}\left(\xi_{2} - \xi_{1},0\right),$}; 
	\node (q3) at (2.5,-0.7) {$q_{3}\left(\xi_{2} - \frac{1}{2} \xi_{1},0\right),$}; 
	\draw[->] (q2)--(1.3,0)--(q1);	\draw[->] (q3)--(1.3,0)--(q1); 
	\node (q4) at (5,0.7) {$q_{4}(\infty,0)$};
	\node (q5) at (7.7,0.7) {$q_{5}(U_{4}=0, V_{4} =0)$};
	\node (q6) at (11.2,0.7) {$q_{6}(U_{5}=0, V_{5} =0)$};
	\draw[->] (q6)--(q5); 	\draw[->] (q5)--(q4);
	\node (q7) at (5,-0.7) {$q_{7}(2,0)$}; \draw[->] (q7)--(5,0)--(11.2,0)--(q6);
	\node (q8) at (7,-0.7) {$q_{8}(-6s,0)$}; \node (q9) at (10.7,-0.7) {$q_{9}(2(-2 + 9 s^{2} + 3\xi_{1} - 4\xi_{2} ),0).$};
	\draw[->] (q8)--(q7); \draw[->] (q9)--(q8);
	\end{tikzpicture}
\end{equation}
and the resulting point configuration and its resolution via blowups is shown on Figure~\ref{fig:vdPT-soic-4}.
We see that there are $-3$ curves and so this surface family is not minimal, but by blowing down the $-1$-curve $H_{y} - F_{1}$ we
can transforms into the $E_{6}^{(1)}$ surface. The standard geometric realization for such surface is given in 
\cite[Section 8.2.22]{KajNouYam:2017:GAPE} and is shown on Figure~\ref{fig:surface-e61}. Here we use $(q,p)$ coordinates for the main affine chart
and the base points configuration, in terms of the root variables $a_{i}$ normalized as $a_{0} + a_{1} + a_{2} = 1$, is
\begin{equation}\label{eq:basept-e61}
p_{1}(\infty,0) \leftarrow p_{2}(0,-a_{2}),\quad p_{3}(0,\infty)\leftarrow p_{4}(a_{1},0),\quad
p_{5}(\infty,\infty) \leftarrow p_{6}\left(0,1\right)\leftarrow p_{7}\left(0,t\right)\leftarrow p_{8}\left(0, a_{0}+t^{2}\right).	
\end{equation}

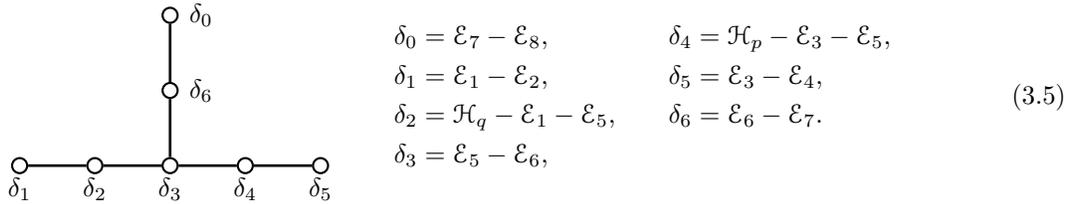
\begin{figure}[ht]
\begin{equation}\label{eq:d-roots-e61}			
	\raisebox{-40pt}{\begin{tikzpicture}[
			elt/.style={circle,draw=black!100,thick, inner sep=0pt,minimum size=2mm}]
		\path 	(-2,0) 	node 	(d1) [elt] {}
		        (-1,0) 	node 	(d2) [elt] {}
		        ( 0,0) 	node  	(d3) [elt] {}
		        ( 1,0) 	node  	(d4) [elt] {}
		        ( 2,0) 	node 	(d5) [elt] {}
		        ( 0,1)	node 	(d6) [elt] {}
		        ( 0,2)	node 	(d0) [elt] {};
		\draw [black,line width=1pt ] (d1) -- (d2) -- (d3) -- (d4) -- (d5)  (d3) -- (d6) -- (d0); 
			\node at ($(d1.south) + (0,-0.2)$) 	{$\delta_{1}$};
			\node at ($(d2.south) + (0,-0.2)$)  {$\delta_{2}$};
			\node at ($(d3.south) + (0,-0.2)$)  {$\delta_{3}$};
			\node at ($(d4.south) + (0,-0.2)$)  {$\delta_{4}$};		
			\node at ($(d5.south) + (0,-0.2)$)  {$\delta_{5}$};		
			\node at ($(d6.east) + (0.3,0)$) 	{$\delta_{6}$};		
			\node at ($(d0.east) + (0.3,0)$) 	{$\delta_{0}$};		
			\end{tikzpicture}} \qquad
			\begin{alignedat}{2}
			\delta_{0} &= \mathcal{E}_{7} - \mathcal{E}_{8}, &\qquad  \delta_{4} &= \mathcal{H}_{p} -  \mathcal{E}_{3} - \mathcal{E}_{5},\\
			\delta_{1} &= \mathcal{E}_{1} - \mathcal{E}_{2}, &\qquad  \delta_{5} &= \mathcal{E}_{3} - \mathcal{E}_{4},\\
			\delta_{2} &= \mathcal{H}_{q} - \mathcal{E}_{1} - \mathcal{E}_{5}, &\qquad  \delta_{6} &= \mathcal{E}_{6} - \mathcal{E}_{7}.\\
			\delta_{3} &= \mathcal{E}_{5} - \mathcal{E}_{6}, 			
			\end{alignedat}
\end{equation}
	\caption{The Surface Root Basis for the standard $E_{6}^{(1)}$ Sakai surface}
	\label{fig:d-roots-e61}	
\end{figure}

The surface root basis for this geometric realization is shown on Figure~\ref{fig:d-roots-e61}	
and the corresponding symmetry root basis is shown on Figure~\ref{fig:a-roots-a21}.

\begin{figure}[ht]
\begin{equation}\label{eq:a-roots-a21}			
	\raisebox{-32.1pt}{\begin{tikzpicture}[
			elt/.style={circle,draw=black!100,thick, inner sep=0pt,minimum size=2mm}]
		\path 	(0,1.73) 	node 	(a0) [elt, label={[xshift=0pt, yshift = 0 pt] $\alpha_{0}$} ] {}
		        (-1,0) node 	(a1) [elt, label={[xshift=-10pt, yshift = -10 pt] $\alpha_{1}$} ] {}
		        ( 1,0) 	node  	(a2) [elt, label={[xshift=10pt, yshift = -10 pt] $\alpha_{2}$} ] {};
\		\draw [black,line width=1pt ] (a0) -- (a1) -- (a2) -- (a0);
	\end{tikzpicture}} \qquad
			\begin{aligned}
			\alpha_{0} &= \mathcal{H}_{q} + \mathcal{H}_{p} - \mathcal{E}_{5} - \mathcal{E}_{6} - \mathcal{E}_{7} - \mathcal{E}_{8}, \\
			\alpha_{1} &= \mathcal{H}_{q} - \mathcal{E}_{3} - \mathcal{E}_{4}, \\
			\alpha_{2} &= \mathcal{H}_{p} - \mathcal{E}_{1} - \mathcal{E}_{2},\\[5pt]
			\delta & = \mathrlap{\alpha_{0} + \alpha_{1} + \alpha_{2}.} 
			\end{aligned}
\end{equation}
	\caption{The Symmetry Root Basis for the standard $E_{6}^{(1)}$ Sakai surface}
	\label{fig:a-roots-a21}	
\end{figure}

Rather than blowing down the curve $H_{y} - F_{1}$ it is more convenient to blow up an additional point $p_{9}$ on the standard configuration that we 
take to be the intersection point of $H_{p} - E_{3} - E_{5}$ and $E_{3} - E_{4}$ as shown on Figure~\ref{fig:surface-e61}, 
i.e., $p_{9}(u_{3} = v_{3} = 0)$. Then we have the following correspondence.

\begin{lemma}\label{lem:IP-to-Ok-4} The change of bases for Picard lattices between the space of initial conditions for the  van der Put--Top 
	surface shown on Figure~\ref{fig:vdPT-soic-4} and the standard $E_{6}^{(1)}$ surface shown on Figure~\ref{fig:surface-e61}
	(with an additional blowup point $p_{9}$) is given by 
	\begin{equation}\label{eq:basis-vdPT-KNY}
		\begin{aligned}
			\mathcal{H}_{x} &= 2\mathcal{H}_{q} + \mathcal{H}_{p} - \mathcal{E}_{1} - \mathcal{E}_{2} - \mathcal{E}_{3} - \mathcal{E}_{9}, &\qquad 
				\mathcal{H}_{q} & = \mathcal{H}_{y},\\
			\mathcal{H}_{y} & = \mathcal{H}_{q},  &\qquad 	
				\mathcal{H}_{p} &= \mathcal{H}_{x} + 2\mathcal{H}_{y} - \mathcal{F}_{1} - \mathcal{F}_{2} - \mathcal{F}_{4} - \mathcal{F}_{5}, \\
			\mathcal{F}_{1}	&= \mathcal{H}_{q} - \mathcal{E}_{9}, &\qquad 
				\mathcal{E}_{1} &= \mathcal{H}_{y} - \mathcal{F}_{5},\\ 
			\mathcal{F}_{2}	&= \mathcal{H}_{q} - \mathcal{E}_{3}, &\qquad 
				\mathcal{E}_{2} &= \mathcal{H}_{y} - \mathcal{F}_{4},\\ 
			\mathcal{F}_{3}	&= \mathcal{E}_{4}, &\qquad 
				\mathcal{E}_{3} &= \mathcal{H}_{y} - \mathcal{F}_{2},\\ 
			\mathcal{F}_{4}	&= \mathcal{H}_{q} - \mathcal{E}_{2}, &\qquad 
				\mathcal{E}_{4} &= \mathcal{F}_{3},\\ 
			\mathcal{F}_{5}	&= \mathcal{H}_{q} - \mathcal{E}_{1}, &\qquad 
				\mathcal{E}_{5} &= \mathcal{F}_{6},\\ 
			\mathcal{F}_{6}	&= \mathcal{E}_{5}, &\qquad 
				\mathcal{E}_{6} &= \mathcal{F}_{7},\\ 
			\mathcal{F}_{7}	&= \mathcal{E}_{6}, &\qquad 
				\mathcal{E}_{7} &= \mathcal{F}_{8},\\ 
			\mathcal{F}_{8}	&= \mathcal{E}_{7}, &\qquad 
				\mathcal{E}_{8} &= \mathcal{F}_{9},\\ 
			\mathcal{F}_{9}	&= \mathcal{E}_{8}, &\qquad 
				\mathcal{E}_{9} &= \mathcal{H}_{y} - \mathcal{F}_{1}.
		\end{aligned}
	\end{equation}
\end{lemma}

Using the symplectic form $k dx\wedge dy$ to compute the root variables, we get 
\begin{equation*}
		a_{0}^{\mathrm{vdPT}} = k\left(1 - \frac{\xi_{1}}{2} + \xi_{2}\right),\quad 
		a_{1}^{\mathrm{vdPT}} =  - k \frac{\xi_{1}}{2},\quad 
		a_{2}^{\mathrm{vdPT}} = k (\xi_{1} - \xi_{2}),				
	\end{equation*}
and the normalization condition $a_{0} + a_{1} + a_{2} = 1$ results in  $k = 1$, so the standard symplectic structure differs by a sign from the form
$dy \wedge dx$ used to obtain equations \eqref{eq:ex9}. This can be fixed by switching the sign in the Hamiltonian \eqref{eq:ex9-Ham}, and so we put
$\symp{vdPT} = dx\wedge dy$ and 
\begin{equation}\label{eq:ex9-Hammod}
	\Ham{vdPT}{IV} = 3 x^{2} y^{3} + \frac{9 \xi_{1}- 12 \xi_{2}}{2} x y^{2} + \frac{3\xi_{1}^{2} - 9 \xi_{1} \xi_{2} + 6 \xi_{2}^{2}}{2} y
		- \frac{3}{2} x - \frac{9 x y}{2}s.
\end{equation}
Equations \eqref{eq:ex9} then can be written as 
\begin{equation}\label{eq:ex9mod}
	\left\{
	\begin{aligned}
		\frac{dx}{ds} &= - \frac{\partial \Ham{vdPT}{IV}}{\partial y} = \frac{3(-\xi_{1} + 2 \xi_{2})(-\xi_{2} + \xi_{1})}{2} + \frac{9 s}{2} x
			+ (-9\xi_{1} + 12\xi_{2}) x y - 9 x^{2} y^{2},\\
		\frac{dy}{ds} &=  \frac{\partial \Ham{vdPT}{IV}}{\partial x} = -\frac{3}{2} + 6 x y^{3} + \frac{9 \xi_{1} - 12 \xi_{2}}{2} y^{2} - \frac{9 y s}{2}.
	\end{aligned}
	\right.
\end{equation}

It is well-known that, up to some scaling, the standard $E_{6}^{(1)}$ surface is the space of initial conditions for the differential 
$\Pain{IV}$ equation 
\begin{equation}\label{eq:P4-std}
\Pain{IV}={\Pain{IV}}_{\alpha,\beta}:\quad \frac{d^2 w}{dt^2} = \frac{1}{2w}\left(\frac{d w}{dt}\right)^2 + 
\frac{3}{2}w^3 + 4tw^2 + 2(t^2 - \alpha)w + \frac{\beta}{w},
\end{equation}
where $t$ is an independent variable, $w=w(t)$ is a dependent variable, and $\alpha$, $\beta$ are complex parameters. The Hamiltonian 
form of this equation is given by Okamoto \cite{Oka:1980:PHAWPE,Oka:1980:PHAWPEIDESPH,Oka:1986:SPEISFPEPP} and it takes the form 
  \begin{equation}\label{eq:Ok-Ham-4}
  	\left\{
 	\begin{aligned}
 		\frac{df}{dt} &= \frac{\partial \Ham{Ok}{IV}}{\partial g} =4fg-f^2-2tf-2\kappa_0,\\
		\frac{dg}{dt} &= -\frac{\partial \Ham{Ok}{IV}}{\partial f} =-2g^2+(2f+2t)g -\theta_{\infty},			
 	\end{aligned}
 	\right.\qquad\text{where}\quad
	\begin{aligned}
	\Ham{Ok}{IV}(f,g;t) &= 2fg^2 - (f^2 + 2tf + 2 \kappa_0)g + \theta_\infty f.	\\
	\symp{Ok} &= dg\wedge df.
	\end{aligned}
  \end{equation}
Eliminating the function $g=g(t)$ from these equations we get $\Pain{IV}$ equation \eqref{eq:P4-std} for the function $f=f(t)$ with parameters 
\begin{equation}\label{eq:Ok-pars-4}
\alpha=1+2\theta_{\infty}-\kappa_0,\quad \beta=-2\kappa_0^2.
\end{equation}
The configuration of the base points for this system is 
\begin{equation}\label{eq:basept-Okamoto-4}
\begin{aligned}
q_{1}(\infty,0) &\leftarrow q_{2}(0,\theta_{\infty}),\qquad q_{3}(0,\infty)\leftarrow q_{4}(\kappa_{0},0),\\
q_{5}(\infty,\infty) &\leftarrow q_{6}(0,2)\leftarrow q_{7}(0,-4t)
\leftarrow q_{8}(0,4(1 + 2 t^{2} + \theta_{\infty} - \kappa_{0})),	
\end{aligned}
\end{equation}
which is almost the same as \eqref{eq:basept-e61}.

Using the linear change of basis \eqref{eq:basis-vdPT-KNY} we can match system \eqref{eq:ex9mod} with the Okamoto system \eqref{eq:Ok-Ham-4}.

\begin{theorem}\label{thm:coords-IP-Ok-4} The change of coordinates and parameter matching between the van der Put--Top space of initial conditions and the 
	Okamoto space of initial conditions for $\Pain{IV}$ is given by 
    \begin{equation*}
   	 \left\{\begin{aligned}
   	 	f(x,y,s) &= - \frac{\mathfrak{i}}{y},\\
   		g(x,y,s) &= - \mathfrak{i} y (xy + \xi_{1} - \xi_{2}),\\
		t(s) &= - \frac{3 \mathfrak{i} s}{2},\\  
		\kappa_{0} &= 1-\frac{\xi_{1}}{2},\\
		\theta_{\infty} &= \zeta_{2} - \zeta_{1},\\
   	 \end{aligned}\right.
    \qquad\text{and conversely} \qquad 
    	\left\{\begin{aligned}
   	 	x(f,g,t) &= - \mathfrak{i} f (q p - \theta_{\infty}),\\
   		y(f,g,t) &= - \frac{\mathfrak{i}}{f},\\
		s(t) &= \frac{2 \mathfrak{i} t}{3},\\  
		\xi_{1} &= - 2 \kappa_{0}\\
		\xi_{2} &= \theta_{\infty} - 2 \kappa_{0}.
    	\end{aligned}\right.
    \end{equation*}
\end{theorem}

To match the Hamiltonians $\Ham{Ok}{IV}$ and \eqref{eq:ex9-Hammod}, we proceed as in the previous example:
\begin{align*}
	\Symp{vdPT} &= \symp{vdPT} - d\Ham{vdPT}{VI}\wedge ds = \varphi^{*}(\Symp{Ok}) =  \varphi^{*}\left(dq\wedge  - d\Ham{Ok}{VI}\wedge dt\right) \\
	&= (- \mathfrak{i} y^{2}) dx \wedge\left(\frac{i}{y^{2}}dy\right) - \left(\frac{-3 \mathfrak{i} }{2}\right)d\Ham{Ok}{VI}\wedge ds = 
	dx\wedge dy - d\Ham{vdPT}{IV} \wedge ds.
\end{align*}
Direct computation shows that 
\begin{equation*}
	\Ham{vdPT}{VI}(x,y,s) = \frac{-3 \mathfrak{i}}{2}\Ham{Ok}{VI}\left(f(x,y,s),g(x,y,s),t(s)\right)  + \frac{9(\xi_{1} - \xi_{2})s}{2}
\end{equation*}
and the function 
\begin{equation*}
	f(t) = \frac{1}{\mathfrak{i} y\left(\frac{2 \mathfrak{i} t}{3}\right)}
\end{equation*}
satisfies equation \eqref{eq:P4-std} with $\alpha = 1 - \frac{3 \xi_{1}}{2} + 2 \xi_{2}$, $\beta = - \frac{\xi_{1}^{2}}{2}$.

\section{Acknowledgments} 
The author is very grateful to Galina Filipuk and, especially, to Alexander Stokes, for many useful discussions. 
\label{sec:acknowledgments}


%
%
%


\bibliographystyle{amsalpha}

\providecommand{\bysame}{\leavevmode\hbox to3em{\hrulefill}\thinspace}
\providecommand{\MR}{\relax\ifhmode\unskip\space\fi MR }
\providecommand{\MRhref}[2]{%
  \href{http://www.ams.org/mathscinet-getitem?mr=#1}{#2}
}
\providecommand{\href}[2]{#2}

\end{document}